\documentclass[%
aps,
prx,
reprint,
superscriptaddress,
nofootinbib,
amsmath,
amssymb
]{revtex4-2}
\usepackage{
    physics,
    graphicx,
    multirow,
    mathtools, 
    placeins, 
    nicefrac, 
    hyperref, 
    threeparttable,
    enumitem,
    booktabs
}

\usepackage[capitalize]{cleveref}
\usepackage{amsmath, amssymb}
\Crefname{section}{Sec.}{Secs.}

\usepackage[dvipsnames]{xcolor}
\hypersetup{
    colorlinks,
    linkcolor={blue!50!black},
    citecolor={blue!50!black},
    urlcolor={blue!80!black}
}
\usepackage{graphicx}
\usepackage[caption=false]{subfig} 
\renewcommand{\var}[1]{{\mathrm{Var}[#1]}}

\newcommand{\mse}[1]{{\mathrm{MSE}[#1]}}
\newcommand{\bias}[1]{{\mathrm{Bias}[#1]}}
\newcommand{\expect}[1]{{\mathrm{E}[#1]}}

\DeclarePairedDelimiterX\pbraket[2]{\langle\!\langle}{\rangle\!\rangle}{#1 \delimsize\vert #2}
\DeclareMathOperator*{\sign}{sgn}

\begin{document}
    
\title{Quantum Error Mitigation for Sampling Algorithms}

\newcommand{\addressPKU}
{\affiliation{Center on Frontiers of Computing Studies, School of Computer Science, Peking University, Beijing 100871, China}}

\newcommand{\addressOxMat}{\affiliation{Department of Materials, University of Oxford, Oxford, OX1 3PH, United Kingdom}}

\newcommand{\addressQM}{\affiliation{Quantum Motion, 9 Sterling Way, London N7 9HJ, United Kingdom}}

\author{Kecheng Liu}
\addressPKU

\author{Zhenyu Cai}\email{cai.zhenyu.physics@gmail.com} \addressOxMat

\date{\today}

\begin{abstract}
Recent experimental breakthroughs have signalled the imminent arrival of the early fault-tolerant era. However, for a considerable period in the foreseeable future, relying solely on quantum error correction for full error suppression will remain extremely challenging due to its substantial hardware overhead. Additional help from quantum error mitigation (QEM) is essential for bridging this gap towards achieving quantum advantage. The application of QEM has so far been restricted to expectation value estimation, leaving its extension to sampling-based algorithms -- which is expected to play a pivotal role in the early fault-tolerant era -- an unresolved challenge. In this work, we present a framework for applying any QEM techniques to obtain the error-mitigated output distribution, showing that this incurs no greater cost than estimating a single observable. We also devised a way to sample from this distribution and constructed an explicit scheme for applying any QEM methods to quantum phase estimation, which can be generalised to other sampling algorithms. Numerical experiments were conducted to validate the efficacy of these methods. We believe our methods significantly broaden the scope of QEM,  extending its applicability to most algorithms of practical interest and forming a crucial step towards realising quantum advantage.
\end{abstract}

\maketitle 

\section{Introduction}
Noise in quantum hardware is the biggest roadblock on the way towards practical implementation of quantum computers. It can be tackled using quantum error corrections (QEC)~\cite{terhalQuantumErrorCorrection2015}, which use additional qubits to encode and protect our quantum information. However, the qubit overhead and additional operations needed for QEC also mean that complete error suppression relying on QEC alone, i.e. full fault tolerance, for large-scale algorithms can still be decades away. As a result, a different error suppression paradigm, quantum error mitigation (QEM)~\cite{caiQuantumErrorMitigation2023}, was developed, which comes with low or sometimes no additional hardware requirements, but relies on extracting information from additional circuit runs instead. 

At the start, QEM was mostly studied in the context of applying directly to physical circuits without QEC~\cite{liEfficientVariationalQuantum2017,temmeErrorMitigationShortDepth2017}, which has also been successfully implemented in many state-of-the-art experiments~\cite{kimScalableErrorMitigation2023,kimEvidenceUtilityQuantum2023,obrienPurificationbasedQuantumError2023,liaoMachineLearningPractical2024}. However, QEM will also be equally applicable to early fault-tolerant machines where the residual logical errors can be addressed using QEM~\cite{suzukiQuantumErrorMitigation2022,piveteauErrorMitigationUniversal2021}. With the recent rapid advance in QEC~\cite{acharyaQuantumErrorCorrection2024,bluvsteinLogicalQuantumProcessor2024,rodriguezExperimentalDemonstrationLogical2024}, it becomes increasingly crucial to study the role of QEM in the early fault-tolerant era. One of the most important problems to address is the applicability of QEM beyond expectation value estimation, which is also a long-standing open problem listed in Ref.~\cite{caiQuantumErrorMitigation2023}. All existing QEM techniques are constructed for algorithms whose output comes from expectation value estimation, e.g. variational algorithm. As we move towards the fault-tolerant era, an increasing number of important algorithms we want to implement are actually based on sampling, e.g. quantum phase estimation. Hence, extending the applicability of QEM to sampling algorithms is critical for achieving quantum advantage in the early fault-tolerant era. Ref.~\cite{suzukiQuantumErrorMitigation2022} has some insightful discussion on adapting QEM to phase estimation using oracles in their appendix, but there is not yet an explicit complete framework for applying QEM to sampling algorithms.

In this paper, we will address this problem by showing that sampling algorithms can actually be efficiently transformed into expectation value estimation problems, for which all existing QEM techniques can be applied. The associated overhead can be analysed by refining the QEM framework defined in Ref. \cite{caiPracticalFrameworkQuantum2021}. In addition, we are also able to see more directly the special role of post-selection QEM techniques compared to the others.

\section{Estimating the error-mitigated distribution}\label{sec:est_distr}
Our target is to estimate the expectation value of some observable $O$ on the ideal state $\rho_0$, but the actual expectation value we manage to obtain is $\Tr(O \rho)$ for some noisy state $\rho$. The goal of quantum error mitigation is to obtain an error-mitigated expectation value closer to the ideal value $\Tr(O\rho_0)$ at the cost of an increased number of circuit runs. Most of the QEM techniques, including probabilistic error cancellation~\cite{temmeErrorMitigationShortDepth2017,endoPracticalQuantumError2018}, linear zero-noise extrapolation~\cite{temmeErrorMitigationShortDepth2017,liEfficientVariationalQuantum2017}, symmetry verification~\cite{mcardleErrorMitigatedDigitalQuantum2019,bonet-monroigLowcostErrorMitigation2018} and virtual purification~\cite{koczorExponentialErrorSuppression2021,hugginsVirtualDistillationQuantum2021,liuVirtualChannelPurification2024}, can be viewed as trying to estimate $\Tr(O \rho_{\mathrm{em}})$ for some error-mitigated state $\rho_{\mathrm{em}}$~\cite{caiPracticalFrameworkQuantum2021} and we will focus on these cases here. It is worth noting that the methods below will be applicable even for QEM methods beyond those described in Ref.~\cite{caiPracticalFrameworkQuantum2021}, but the analytical bounds obtained may vary. 

We observe that all the QEM techniques can be extended to sampling problems by setting the observable $O$ to be the projector $\Pi_z = \ketbra{z}$ for some output string $z$. With $O = \Pi_z$ for different $z$, the corresponding expectation values turn into the output distribution for the corresponding state, and the process of QEM becomes trying to obtain an error-mitigated distribution $\Tr(\Pi_z \rho_{\mathrm{em}}) = p_{\mathrm{em}}(z)$ that can better approximate the ideal distribution $\Tr(\Pi_z \rho_0) = p_0(z)$ than the noisy distribution $\Tr(\Pi_z \rho) = p(z)$.

Following the framework outlined in Ref.~\cite{caiPracticalFrameworkQuantum2021}, which is summarised in \cref{sec:previous_framework}, to estimate the error-mitigated expectation value $p_{\mathrm{em}}(z) = \Tr(\Pi_z \rho_{\mathrm{em}})$, we need to run different variants of the unmitigated circuits (called \emph{response circuits}) with added qubits, gates and/or measurements, and then perform the target measurement $\Pi_z$ at the end, outputting $0$ or $1$. In addition, we need to attach a sign to the measurement outcome based on the circuit configuration that we are running (e.g. linear zero-noise extrapolation, probabilistic error cancellation) and/or the measurement results of additional observables (e.g. virtual purification, symmetry verification). Hence, the output is a random variable $\hat{R}_z$ (called the \emph{response variable}) that takes the value $\pm 1$ and $0$ with the probability $r_{z,\pm}$ and $r_{z,0}$, respectively. In practice, we simply measure the computational basis in every run of the response circuit. The output string $z'$ can be post-processed to obtain a sample of $\Pi_{z}$ measurements for all possible $z$: $\hat{\Pi}_{z'} = 1$ and $\hat{\Pi}_{z} = 0 \quad \forall z \neq z'$. We can further define $r_z = r_{z,+} + r_{z,-} = 1 - r_{z,0}$, which is essentially the probability of obtaining the string $z$ as the output of the ensemble of response circuits.

By taking the average of $\hat{R}_{z}$ over $N_{\mathrm{cir}}$ circuit runs to obtain the sampled mean estimator $\overline{R}_{z}$ and multiplying by a normalising factor $A$, we obtain the error-mitigated estimator $\hat{p}_{\mathrm{em}}(z)$ for the probability of $z$: $\hat{p}_{\mathrm{em}}(z) = A \overline{R}_{z}$. In most cases, we have $A = (p_+ - p_-)^{-1}$ as derived in \cref{sec:previous_framework}, where $p_{\pm}$ are the probabilities of attaching a plus/minus sign to a circuit run. The expectation and variance of this estimator are derived in \cref{sec:prop_distr_est}:
\begin{align}
    \expect{\hat{p}_{\mathrm{em}}(z)}  &= A (r_{z,+} - r_{z,-} ) = p_{\mathrm{em}}(z)\label{eqn:em_distr_expect}\\
    \var{\hat{p}_{\mathrm{em}}(z)} &= \frac{1}{N_{\mathrm{cir}}} \left[A^2r_z - p_{\mathrm{em}}^2(z)\right] \approx \frac{A^2 r_z}{N_{\mathrm{cir}}} \label{eqn:em_distr_var}
\end{align}
The total variance of the estimator of the whole distribution summed over $z$ is bounded by $\frac{A^2 - 1}{N_{\mathrm{cir}}} \leq \sum_{z} \var{\hat{p}_{\mathrm{em}}(z)} \leq \frac{A^2}{N_{\mathrm{cir}}}$. In contrast with the usual relatively loose sampling overhead bound obtained for QEM on general observable using Hoeffding's inequality~\cite{caiQuantumErrorMitigation2023}, what we have here are very tight bounds on $\sum_{z} \var{\hat{p}_{\mathrm{em}}(z)}$ since we usually have $A^2 \gg 1$, which will enable us to obtain very tight bounds on the sampling overhead. Hence, for practical purposes, the total variance of our distribution estimator is:
\begin{align}\label{eqn:tot_em_variance}
   \sum_{z} \var{\hat{p}_{\mathrm{em}}(z)} \approx \frac{A^2}{N_{\mathrm{cir}}}.
\end{align}
Let us look back at the total variance of the naive estimator for the noisy distribution without QEM:
\begin{align*}
    \sum_{z} \var{\hat{p}(z)} = \sum_{z} \frac{p(z) - p^2(z)}{N_{\mathrm{cir}}} =  \frac{1 - \sum_{z} p^2(z)}{N_{\mathrm{cir}}}
\end{align*}
The noisy distribution $p(z)$ is usually quite spread out due to the noise (unless the noise level is very low), which means $\sum_{z} p^2(z) \ll 1$ and we have $\sum_{z} \var{\hat{p}(z)} \approx N_{\mathrm{cir}}^{-1}$. 
Compared to \cref{eqn:tot_em_variance}, we see that the sampling overhead is $A^2$, which is similar to what is needed for estimating one observable before~\cite{caiPracticalFrameworkQuantum2021}. i.e. \emph{the sampling overhead needed for obtaining the error-mitigated distribution is similar to that for estimating one observable.} By using $I  = \sum_{z} \Pi_z$, one can also show that the total variance above is the same as the variance of estimating the identity observable $I$. In \cref{sec:mse_err}, we further discuss the effect of having biases in the estimator, which, just like in other QEM applications, will add to the variance to give us the total square error $\mathrm{TSE}(\hat{p}_{\mathrm{em}}) = \sum_{z} (\hat{p}_{\mathrm{em}}(z) - p_0(z))^2$. 
The total variation distance of the error-mitigated distribution is upper-bounded by $2^{N/2-1}\sqrt{\mathrm{TSE}(\hat{p}_{\mathrm{em}})}$ (see \cref{sec:mse_err}). We get close to this upper bound only when the distributions are close to uniform, thus it should not be a problem for many distributions of practical interests.

\section{Sampling from the error-mitigated distribution}\label{sec:samp_distr}
More generally, our method for estimating the error-mitigated distribution can be transformed into a modified sampling algorithm, in which instead of steadily adding a sample after each circuit run, we have the additional option of removing an existing sample after a circuit run. Thus, the number of effective samples is smaller than the number of circuit runs. In the usual sampling algorithm, if we measure the outcome $z$, we will add a sample into the data bin corresponding to $z$.
In the error-mitigated sampling, algorithm, when we run the ensemble of the response circuit, the sample we obtain is a tuple  $(z, s)$ where $s = \pm 1$ is the sign for the circuit run. We will add one sample to the $z$ bin if $s = +1$ and remove one sample if $s = -1$. Suppose after $N_{\mathrm{cir}}$ circuit runs, the number of runs that output string $z$ is $N_z$, within which $N_{z, \pm}$ are the number of runs that come with the $\pm$ sign. The number of effective samples we obtain for string $z$ is $N_{z,\mathrm{em}} = N_{z, +} - N_{z, -}$.
In the limit of an infinite number of circuit runs, our sampling algorithm here will recover the error-mitigated distribution: $\lim_{N_{\mathrm{cir}} \rightarrow \infty} N_{z,\mathrm{em}} =  N_{\mathrm{cir}} (r_{z,+} - r_{z,-}) \propto p_{\mathrm{em}}(z)$. 
The number of effective samples we obtained after $N_{\mathrm{cir}}$ circuit runs is roughly
\begin{align*}
    N_{\mathrm{samp}} = \sum_{z} N_{z,\mathrm{em}} \approx N_{\mathrm{cir}} \sum_{z}(r_{z, +} - r_{z, -}) = A^{-1}N_{\mathrm{cir}}
\end{align*}
where we have used \cref{eqn:em_distr_expect}. 
The estimator of the distribution obtained by running our new sampling method is $\hat{p}_{\mathrm{em}}'(z) = \frac{N_{z,\mathrm{em}}}{N_{\mathrm{samp}}}$, which is slightly different from the direct estimation $\hat{p}_{\mathrm{em}}(z) = \frac{N_{z,\mathrm{em}}}{A^{-1}N_{\mathrm{cir}}}$ in \cref{sec:est_distr}. 
The new estimator $\hat{p}_{\mathrm{em}}'(z)$ always ensures the normalisation of the whole distribution ($\sum_z\hat{p}_{\mathrm{em}}'(z) = 1$), but in the process introduces additional shot noise in the denominator of the expression. Hence, which one to use depends on the final usage of the distribution. 

One peculiarity of our sampling algorithm is we can have a negative number of samples $N_{z,\mathrm{em}} < 0$ for a given output $z$ due to the option to remove samples. Such a negative value is not physical and is entirely due to the shot noise in the estimator. This can happen when $p_{\mathrm{em}}(z) \ll 1$ and the number of circuit runs is small. Hence, by setting these negative $N_{z,\mathrm{em}}$ to $0$, we are effectively removing shot noise from the estimator and moving closer to $p_{\mathrm{em}}(z)$. We can perform a similar procedure in \cref{sec:est_distr} by setting the negative $\hat{p}_{\mathrm{em}}(z)$ to $0$ followed by renormalisation to further improve our estimation. However, if we want to further increase the number of circuit runs to improve our experiments, we need to restore these negative values before continuing since they are useful for cancelling out the positive-value shot noise later. 

If instead of just obtaining the ground energy, we also want to perform additional measurements on the associated output ground state, then QEM techniques are still applicable in a similar way with additional measurements and post-processing outlined in \cref{sec:eigenstate}.

\section{Post-selection}
One particular class of error suppression technique that goes beyond our discussion so far is post-selection, which appears in quantum error detection and QEM schemes like symmetry verification~\cite{mcardleErrorMitigatedDigitalQuantum2019,bonet-monroigLowcostErrorMitigation2018,mccleanDecodingQuantumErrors2020}. They are natively compatible with sampling problems without needing further modification. The key difference is while post-selection is equivalent to attaching a factor of $0$ or $1$ to the circuit runs, for the other QEM methods described in \cref{sec:est_distr} we are attaching factors of $\pm 1$. Exactly because we are not attaching any negative factor to the circuit runs in post-selection, the output results can be viewed as a valid mixture of the various circuit runs without negative probability. Hence, we are able to obtain a physical error-mitigated state via post-selection instead of a virtual one in post-processing. As a result, any further operation on the output state, including sampling, can be straightforwardly applied. We can broaden our framework to include post-selection by allowing the attachment of the factor $0$, which is discussed in detail in \cref{sec:post-select}. Translating into the sampling algorithm in \cref{sec:samp_distr}, the prefactors of $\pm 1$ and $0$ correspond to adding/removing a sample from the data bin and leaving the bin unchanged, respectively.

Some algorithms come with a natural set of conditions for efficiently verifying their output, e.g. Shor's algorithm. These conditions can then be used as post-selection conditions as well. In the case of Shor's algorithm, it has the perfect post-selection condition since only the correct result can factorise the given number and pass the verification. In this case, there is no stronger post-selection and, more generally, QEM schemes for this algorithm, so there is no need to apply additional QEM on top of Shor's algorithm.

\section{Beyond Recovering the Distribution} \label{sec:sm_str_from_distr}
Let us suppose that instead of trying to obtain the whole output distribution or to sample from it, we are asking more specific questions for which additional post-processing is required. In this section, we will use the example of ground-state-energy estimation using quantum phase estimation, in which we are interested in the smallest string in the output distribution. In order to discuss the efficiency of the additional post-processing steps for the given application, rather than the efficiency inherent to the applied QEM technique, we will assume that the QEM method used is powerful enough such that the same smallest string from $p_{\mathrm{em}}(z)$ and $p_0(z)$ are the same and occur with similar probabilities. This is guaranteed to be the case for bias-free methods like probabilistic error cancellation.

After applying error mitigation to obtain the distribution estimator $\hat{p}_{\mathrm{em}}(z)$, we cannot directly output its smallest string as the answer since shot noise might turn some of the zero-probability entry to non-zero. We need to carry out additional steps to determine whether the probability of a string is truly zero or not. For a given string $z$, the probability estimator $\hat{p}_{\mathrm{em}}(z)$ is some random variable that centers around the true value $p_{\mathrm{em}}(z)$, which can be $0$ or some other positive value. Its variance is given in \cref{eqn:em_distr_var}, and the corresponding standard deviation can be estimated using the samples we obtained:
\begin{align}\label{eqn:var_pem}
    \hat{\sigma}[\hat{p}_{\mathrm{em}}(z)] &\approx \sqrt{\frac{A^2\hat{r}_z}{N_{\mathrm{cir}}}} \approx \epsilon\sqrt{\hat{r}_z}.
\end{align}
Here $\hat{r}_z$ is the sample estimator for $r_z$, and $\epsilon^2$ is the total variance of the error-mitigated estimator given in \cref{eqn:tot_em_variance}. 

The simplest way forward is to perform hypothesis testing with $p_{\mathrm{em}}(z) = 0$ as the null hypothesis. We will set some threshold value $p_{\mathrm{th}}(z)$ for the given sample of $\hat{p}_{\mathrm{em}}(z)$ we obtain such that
\begin{align}\label{eqn:select_hypo}
    \hat{p}_{\mathrm{em}}(z) \leq p_{\mathrm{th}}(z) &\Rightarrow \text{Accept Null: } p_{\mathrm{em}}(z) = 0,\\
    \hat{p}_{\mathrm{em}}(z) > p_{\mathrm{th}}(z) &\Rightarrow \text{Reject Null: } p_{\mathrm{em}}(z) = p_{\mathrm{alt}}(z) > 0.\nonumber
\end{align}
The possible choice of $p_{\mathrm{th}}(z)$ is discussed in \cref{sec:hyp_test}. Without any prior information, the threshold $p_{\mathrm{th}}(z)$ can be set proportional to the standard deviation in \cref{eqn:var_pem} to bound the type I error (false rejection of the null hypothesis). 
In the usual problem setting, we often know that the probability of obtaining the smallest string from the ideal circuit is at least $p_{0}(z_{\mathrm{min}}) \geq p_{\mathrm{b}}$, such that we know that we can obtain the answer in $\order{1/p_{\mathrm{b}}}$ ideal circuit runs. In this case, a good choice of the threshold is $p_{\mathrm{th}}(z) = p_{\mathrm{b}}/2$ and we are able to provide bound for both type I and II errors. This choice will lead to larger type I errors than type II as discussed in \cref{sec:hyp_test}.

The arguments above can be similarly extended to other algorithms that output some string based on the ideal distribution $p_0(z)$ (e.g. the maximum string, the string with the highest probability, etc). Suppose that we are given a QEM method that can output an error-mitigated distribution $p_{\mathrm{em}}(z)$ that can correctly answer the given question (which will always be true for all bias-free QEM methods), then the only thing keeping us from getting the right answer is the shot noise in $\hat{p}_{\mathrm{em}}(z)$. One can gauge the expected magnitude of this shot noise by calculating the sampled standard deviation in \cref{eqn:var_pem} and try to mitigate their effects by designing the right set of hypothesis tests like what we did in this section.

Returning back to the smallest-string problem, one can actually design algorithms that directly output the answer without explicitly obtaining the entire output distribution. In the appendix of Ref.~\cite{suzukiQuantumErrorMitigation2022}, they discuss a way to do this using an oracle that can calculate certain decision problems on the output state for transforming the expectation values into decision bits. In \cref{sec:sm_str_alt}, we present a more explicit protocol without referring to any oracles. There, we can perform hypothesis testing digit-by-digit rather than string-by-string, and thus can achieve advantages over the general method above if one has access to prior information about some marginal probability of the digits of the smallest string.

\begin{figure*}[htbp]
    \subfloat[Ideal $p_{0}(z)$\label{fig:ideal_distr}]{\includegraphics[width=0.33\textwidth]{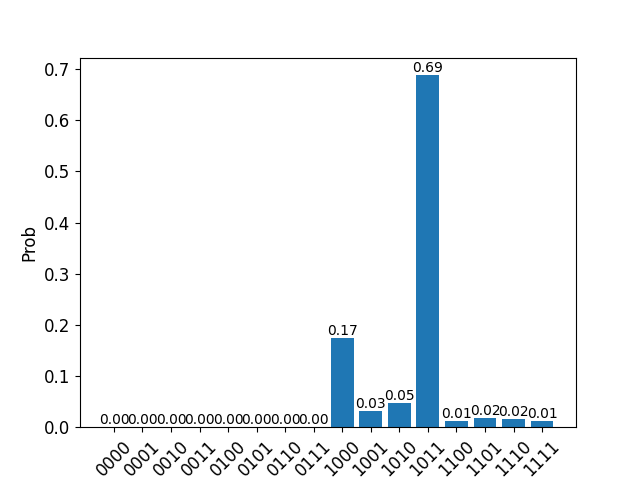}}
    \subfloat[Noisy $p(z)$\label{fig:noisy_distr}]{\includegraphics[width=0.33\textwidth ]{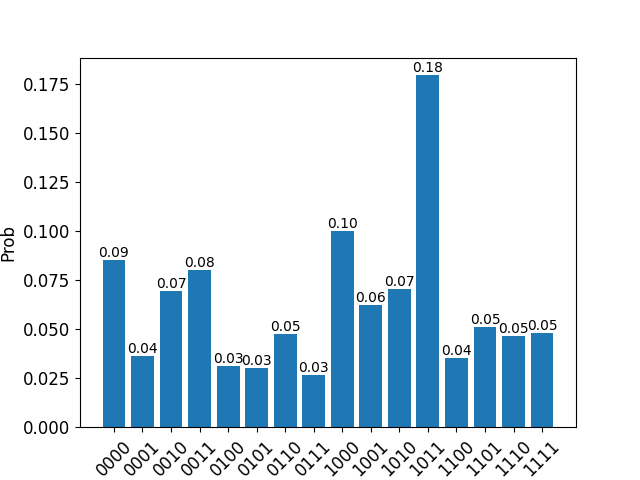}}
    \subfloat[Mitigated $\hat{p}_{\mathrm{em}}(z)$, $10^6$ circuit runs\label{fig:mitigated_distr}]{\includegraphics[width=0.33\textwidth ]{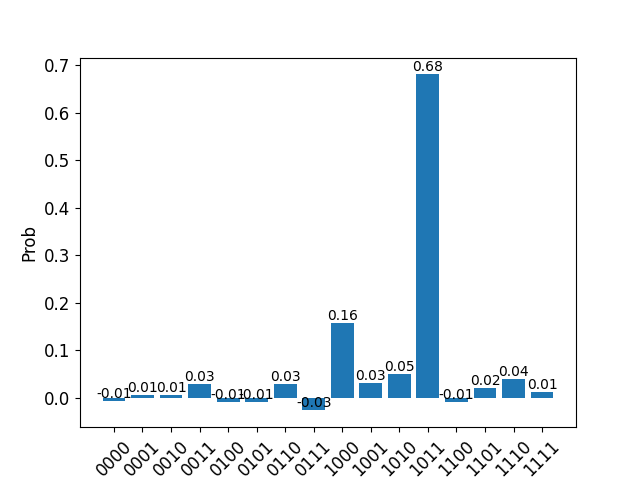}}
    \caption{Output distribution from 4-qubit experiments.}
    \label{fig:distr_recover}
\end{figure*}

\section{Numerical Simulation}\label{sec:numerical_simulation}
In this section, we will be looking at an example of quantum phase estimation circuits modified from the circuit in \cite{QiskittextbookContentChalgorithms}, with further details about the simulation outlined in \cref{sec:further_numerics}. The ideal output distribution from the noiseless circuit of a 4-qubit simulation is shown in \cref{fig:ideal_distr}, in which the smallest string is $1000$. Now let us assume the two-qubit gates in the inverse Fourier transform circuit suffer from depolarising noise such that
the average number of faults per circuit run (circuit fault rate), which is the product of the number of noisy gate $M$ and the per gate error rate $p$, is given by $Mp = 0.6$. As discussed in \cite{caiQuantumErrorMitigation2023}, $\order{1}$ circuit fault rate is the regime where we want to be working in as the gate error rate continues to decrease and enable larger circuit size. Hence, by working with $\order{1}$ circuit fault rate here in our simulation, our small qubit experiment here should be indicative of the performance of larger circuits with similar circuit fault rates. The resultant noise distribution is shown in \cref{fig:noisy_distr}. The total square error of the whole distribution in this case is $\mathrm{TSE}_{\mathrm{noi}} =\sum_{z} (p_0 (z) - p(z))^2 = {0.297}$ and we cannot read off the smallest string from the noisy distribution anymore. 

We will then apply probabilistic error cancellation~\cite{temmeErrorMitigationShortDepth2017,endoPracticalQuantumError2018} on our noisy circuit and obtain samples from $N_{\mathrm{cir}}=10^6$ circuit runs, which outputs the error-mitigated distribution in \cref{fig:mitigated_distr}. We can immediately see that it is much closer to the ideal distribution, with the total square error reduced from $\mathrm{TSE}_{\mathrm{noi}} = {0.297
}$ to $\mathrm{TSE}_{\mathrm{em}} =\sum_{z} (p_0 (z) - \hat{p}_{\mathrm{em}}(z))^2 ={0.004}$. Using the hypothesis testing method outlined in \cref{sec:sm_str_from_distr}, if we set the threshold to be between $0.03 \leq p_{\mathrm{th}} < 0.16$, then the probability of all strings below $1000$ will be set to $0$ and we will recover $1000$ to be the smallest string as before. As mentioned before, in practice, we often know that the probability of obtaining the smallest string is lower bounded by $p_{b}$ for which we can choose the threshold to be $p_b/2$. Hence, we can obtain the right result in this case if we know beforehand the ground state energy will appear with a probability greater than $0.03 \times 2 = 0.06$. Further simulation results on fewer circuit runs and more qubits can be found in \cref{sec:further_numerics}. For bias-free QEM, the average total square error is simply the total variance in \cref{eqn:tot_em_variance} and thus will be inversely proportional to the number of circuit runs. As long as we are operating at the same $\order{1}$ circuit fault rates, the sampling overhead and the mitigation power achieved should be similar even if we increase the number of qubits~\cite{caiQuantumErrorMitigation2023}.

\section{Discussion and Outlook} 
For the task of obtaining the output distribution of a given quantum circuit, by viewing it as measuring the projection operators for all binary outputs -- achieved through post-processing on measurements in the computational basis -- we enable the application of any QEM methods to this task. Using the theoretical framework of linear QEM~\cite{caiPracticalFrameworkQuantum2021}, we analytically derive the sampling overhead for estimating the output distribution, showing that it is comparable to that of estimating a single observable. Hence, for any noisy quantum state whose error-mitigated observable can be efficiently estimated, we can also efficiently obtain its error-mitigated output distribution. 

We further propose an algorithm to sample from the error-mitigated distribution by allowing for the removal of samples, which comes with a similar overhead as estimating the distribution. In the process, we also formalise relations between post-selection QEM and other QEM, highlighting the key difference as whether a negative factor is attached to the circuit output. This distinction provides a basis for analysing differences in applicability and performance. We then show that by combining these protocols with hypothesis testing, we can apply QEM to any sampling algorithms. We outline the explicit protocols for quantum phase estimation, which is further verified in our numerics. 

With this, we believe we have extended QEM to most algorithms of practical interest. Of course, this does not imply that the application of QEM is necessarily efficient in all of these contexts. Its applicability and efficiency remain subject to the same constraints as applying QEM to expectation value estimation, which depends on factors such as the amount of noise in the circuits, prior knowledge about the noise and states, etc~\cite{caiQuantumErrorMitigation2023}. Future work could focus on improving our schemes by exploring more efficient hypothesis testing methods and more tailored ways to apply QEM for specific algorithms similar to the scheme in \cref{sec:sm_str_alt}. It will also be interesting to study the possible interplay between QEC and QEM in the context of sampling algorithms.

\section*{Acknowledgements}
The authors would like to thank the anonymous reviewers in QCTiP for the helpful comments on including the comparison to total variation distance. 
ZC acknowledges support from the EPSRC QCS Hub EP/T001062/1, EPSRC projects Robust and Reliable Quantum Computing (RoaRQ, EP/W032635/1), Software Enabling Early Quantum Advantage (SEEQA, EP/Y004655/1) and the Junior Research Fellowship from St John’s College, Oxford.

\appendix
\section{Derivation for error-mitigated distributions}
\subsection{Recap of the linear QEM framework}\label{sec:previous_framework}
In Ref.~\cite{caiPracticalFrameworkQuantum2021}, the essential idea is a lot of QEM methods can be reframed as trying to prepare the effective error-mitigated state $\rho_{\mathrm{em}}$ and the error-mitigated expectation value is simply obtained by measuring the observable of interests on this error mitigated state $ \expect{\hat{O}_{\mathrm{em}}}= \Tr(O\rho_{\mathrm{em}})$. The error-mitigated state is prepared as the linear combination of the output states of an ensemble of circuits that we call the response circuits:
\begin{align*}
    \rho_{\mathrm{em}} = \sum_{i} \alpha_i \rho_{\mathrm{rsp},i}
\end{align*}
where $\rho_{\mathrm{rsp},i}$ is the output state for the $i$th response circuit, which can include post-selection based on the measurement outcome of some additional auxiliary observables. Here $\alpha_i$ is the coefficient for the corresponding response circuit, and if the response circuit includes post-selection, then the corresponding success probability is absorbed in $\alpha_i$ such that $\rho_{\mathrm{rsp},i}$ is normalized. Note that this implies $\alpha_i$ and in particular its sign can be different for different circuit configurations and different measurement outcomes for the auxiliary observables. 

An efficient way to build an estimator of $\Tr(O\rho_{\mathrm{em}})$ is to use the Monte-Carlo method, which requires us to transform the above equation into
\begin{align*}
    \expect{\hat{O}_{\mathrm{em}}} = \Tr(O\rho_{\mathrm{em}}) = A \sum_{i} \sign(\alpha_i) \frac{\abs{\alpha_i}}{A} \Tr(O\rho_{\mathrm{rsp},i})
\end{align*}
with $A = \sum_{i} \abs{\alpha_i}$.
Hence, in the Monte-Carlo method, the circuit and auxiliary observable outcome that output $\rho_{\mathrm{rsp},i}$ will be sampled with $\frac{\abs{\alpha_i}}{A}$ probability, we will then measure $O$ on this output state and attach the sign $\sign(\alpha_i)$ to the outcome, obtaining an estimator that we denote as $\hat{R}$. The error mitigated estimator is simply obtained by multiplying $\hat{R}$ by a normalisation factor $A$:
\begin{align*}
    \expect{\hat{R}} &=  \sum_{i} \sign(\alpha_i) \frac{\abs{\alpha_i}}{A} \Tr(O\rho_{\mathrm{rsp},i})\\
    \hat{O}_{\mathrm{em}} & = A\hat{R}.
\end{align*}
Note that in most QEM methods, $ \rho_{\mathrm{em}}$ is normalised ($\Tr(\rho_{\mathrm{em}}) = 1$), which implies $\sum_{i} \alpha_i = 1$. In this case, we have
\begin{align*}
    A = (p_+ - p_-)^{-1} \quad \text{with } p_{\pm} = \sum_{\sign(\alpha_i) = \pm 1}  \frac{\abs{\alpha_i}}{A}
\end{align*}
i.e. they are the probability of $\alpha_i$ taking the plus and minus signs, respectively.

\subsection{Derivation of Properties of the Distribution Estimator}\label{sec:prop_distr_est}
The expectation value and the variance of the response variable in \cref{sec:est_distr} are:
\begin{align*}
    \expect{\hat{R}_{z}} &= r_{z,+} - r_{z,-}\\
    \var{\hat{R}_{z}}  &= \expect{\hat{R}_{z}^2} - \expect{\hat{R}_{z}}^2 = (r_{z,+} + r_{z,-}) - (r_{z,+} - r_{z,-})^2
\end{align*}
Hence, the expectation value and variance of $\hat{p}_{\mathrm{em}}(z) = A \overline{R}_{z}$ is given by
\begin{align}
    \expect{\hat{p}_{\mathrm{em}}(z)}  &=  A \expect{\overline{R}_{z}} = A (r_{z,+} - r_{z,-} )\label{eqn:em_distr_expect_app}\\
    \var{\hat{p}_{\mathrm{em}}(z)}  &= \frac{A^2}{N_{\mathrm{cir}}} \var{\hat{R}_{z}} \\
    &= \frac{A^2}{N_{\mathrm{cir}}} \left[(r_{z,+} + r_{z,-}) - (r_{z,+} - r_{z,-})^2\right].
\end{align}
Further using \cref{eqn:em_distr_expect_app}, we have:
\begin{align}\label{eqn:em_distr_var_app}
    \var{\hat{p}_{\mathrm{em}}(z)}  &= \frac{1}{N_{\mathrm{cir}}} \left(A^2 r_z - p_{\mathrm{em}}^2(z)\right) \approx \frac{A^2 r_z}{N_{\mathrm{cir}}} 
\end{align}
where the approximation uses the fact that $Ar_z = p_{\mathrm{em}}(z) + 2Ar_{z,-} \geq p_{\mathrm{em}}(z) \geq p_{\mathrm{em}}^2(z)$ and also the fact that we usually have $A \gg 1$.

We can obtain the total variance of the estimator of the whole distribution by summing over all $z$:
\begin{align*}
    \sum_{z} \var{\hat{p}_{\mathrm{em}}(z)}  &= \frac{1}{N_{\mathrm{cir}}} \left(A^2 - \sum_z p_{\mathrm{em}}^2(z)\right).
\end{align*}
which means 
\begin{align*}
    \frac{A^2 - 1}{N_{\mathrm{cir}}} \leq \sum_{z} \var{\hat{p}_{\mathrm{em}}(z)} \leq \frac{A^2}{N_{\mathrm{cir}}}
\end{align*}

\subsection{Total Mean Square Error of the Distribution}\label{sec:mse_err}
In \cref{sec:est_distr}, we have only talked about shot noise, i.e. how different our sample estimate $\hat{p}_{\mathrm{em}}(z)$ from the error-mitigated distribution $p_{\mathrm{em}}(z)$. Another source of error comes from the difference between the error-mitigated distribution $p_{\mathrm{em}}(z)$ and the ideal distribution $p_0(z)$, which is quantified by the bias of the estimator. For a given $z$, the bias is given by
\begin{align*}
    \bias{\hat{p}_{\mathrm{em}}(z)} = p_{\mathrm{em}}(z) - p_0(z).
\end{align*}
which can be used to obtain the total variational distance between $p_{\mathrm{em}}$ and $p_0$
\begin{align*}
   \delta(p_{\mathrm{em}}, p_0)= \frac{1}{2} \sum_{z} \abs{\bias{\hat{p}_{\mathrm{em}}(z)}}.
\end{align*}
In some QEM methods, like probabilistic error cancellation with perfectly characterized gate noise, it is possible to achieve almost zero bias. Another quantity of interest will be the sum of squares of biases, which can be bound by
\begin{align}\label{eqn:bias_bound}
\sum_{z} \bias{\hat{p}_{\mathrm{em}}(z)}^2 \leq 2\delta(p_{\mathrm{em}}, p_0) \leq 2.
\end{align}
using the fact that $\bias{\hat{p}_{\mathrm{em}}(z)} \leq 1$ and $\delta(p_{\mathrm{em}}, p_0) \leq 1$.

The full error, that is the expected (average-case) square deviation of our estimator $\hat{p}_{\mathrm{em}}(z)$ from the true value $p_0(z)$, for a given $z$ can be quantified using the mean square errors
\begin{align*}
    \mse{\hat{p}_{\mathrm{em}}(z)} &= (\hat{p}_{\mathrm{em}}(z) - p_0(z))^2 \\
    &= \var{\hat{p}_{\mathrm{em}}(z)} + \bias{\hat{p}_{\mathrm{em}}(z)}^2
\end{align*}
and the total mean square error is simply
\begin{align*}
    \mathrm{TSE}(\hat{p}_{\mathrm{em}}) &=  \sum_z \mse{\hat{p}_{\mathrm{em}}(z)} \\
    &\approx \frac{A^2}{N_{\mathrm{cir}}} + \sum_z \bias{\hat{p}_{\mathrm{em}}(z)}^2
\end{align*}
where we have used \cref{eqn:tot_em_variance} and the bias term can be upperbound using \cref{eqn:bias_bound}.

Our use of total square errors here (and similarly the total variance) follows naturally from the estimation of observable in regular QEM, but deviates from the standard distance metrics for distributions, which is the total variation distance $\mathrm{TVD}(\hat{p}_{\mathrm{em}}) = \frac{1}{2}\sum_{z} \abs{\hat{p}_{\mathrm{em}}(z) - p_0(z)}$. Using the usual inequality between $1$-norm and $2$-norm, we have:
\begin{align}
    \frac{1}{2}\sqrt{\mathrm{TSE}(\hat{p}_{\mathrm{em}})} \leq \mathrm{TVD}(\hat{p}_{\mathrm{em}}) \leq \frac{1}{2}2^{N/2}\sqrt{\mathrm{TSE}(\hat{p}_{\mathrm{em}})}
\end{align}
where $N$ is the length of the output string $z$. Hence, in the worst case, $\mathrm{TVD}(\hat{p}_{\mathrm{em}}) $ can be exponentially larger than $\sqrt{\mathrm{TSE}(\hat{p}_{\mathrm{em}})}$. However, this upper bound will only be saturated, when the errors $\abs{\hat{p}_{\mathrm{em}}(z) - p_0(z)}$ are uniformly distributed over all entries. This will only be likely for a uniform $p_{\mathrm{em}}(z)$ if considering only the shot noise in the entries. When considering only the bias in the entries, it is almost impossible to have an error-mitigated distribution that deviates from the ideal distribution with the same magnitude in all entries. Hence, it is unlikely that we will get close to this upper bound for distributions of practical interests.

\section{Post-Selection and Post-Processing}\label{sec:post-select}
Even though technically post-selection is also a type of post-processing, in this section we will use post-processing to specifically refer to the QEM techniques described in \cref{sec:est_distr}, not including post-selection. In post-selection, we know that the ideal output state must satisfy certain constraints and we can effectively measure this constraint operator $\Pi_S$ which outputs $1$ if the constraint is satisfied and $0$ if not. This gives us a simple scheme of mitigating errors by post-selecting the circuit runs by measuring $\Pi_S$ on the output state and keeping only the circuit runs that output $1$. We can perform any additional measurements needed on these post-selected circuit runs, including performing sampling algorithms, to directly yield the error-mitigated samples/expectation values without further post-processing. In post-processing, we are attaching $\pm 1$ factor to each circuit run as discussed in \cref{sec:est_distr}. On the other hand, the key difference for post-selection is that we do not attach any negative factors, instead, we are essentially assigning a $0$ pre-factor to the circuit runs that get filtered out. This translates into $r_{z,-} = 0$ and $A = p_{+}$ for the discussion in \cref{sec:est_distr}. 

However, in many cases, the constraint operator $\Pi_S$ is a global operator that is hard to measure directly, and instead, it is obtained via local measurements and post-processing. Even though a correct value of $\Pi_S$ can be obtained via such measurements, the resultant output state is not the target state anymore and we cannot measure any further observables that we want. Suppose the target observable is $O$, then the error-mitigated expectation value will require the measurement of $\Pi_SO$. In the case where $\Pi_S$ is the projector for Pauli symmetry, it can be hard to measure $\Pi_SO$ directly, but one can measure $(1-\Pi_S)O$ instead, which can be combined with the expectation value of $O$ to obtain the expectation value of $\Pi_SO$ via post-processing~\cite{bonet-monroigLowcostErrorMitigation2018}. In a similar fashion, we can often effectively obtain the post-selected expectation value using a linear combination of other measurements via post-processing. Such a post-processing approach can be applied to sampling problems using the framework outlined in \cref{sec:est_distr}.

\section{Alternative Method for obtaining smallest string}\label{sec:sm_str_alt}
Suppose the ideal output state is $\rho_0$, on which we can perform measurements to obtain a binary string of length $N$. We want to know the smallest string out of all possible outputs. We will use $I_{n}$ to denote identity acting on $n$ qubits and $p_0(y_1y_2...y_k)$ to denote the marginal probability of the \emph{first $k$ bits} being $y_1y_2...y_k$. 

We will use $z_{\mathrm{min}} = z_1z_2...z_N$ to denote the smallest string and its digits. Starting from $z_1$ onwards, the $k$th digit of this string can be obtained by measuring $\ketbra{z_1z_2...z_{k-1}0} \otimes I_{N-k}$ on $\rho_0$
\begin{align*}
    &\Tr(\ketbra{z_1...z_{k-1}0} \rho_0) = p_0(z_1...z_{k-1} 0) = 0\\ 
    &\quad \Rightarrow \quad z_k = 1;\\
    &\Tr(\ketbra{z_1...z_{k-1}0} \rho_0) = p_0(z_1...z_{k-1} 0) > 0\\
    &\quad \Rightarrow \quad z_k = 0.
\end{align*}

Now if noise corrupts the output state, we can still apply the same protocol but with each expectation value above estimated using QEM. The net effect is essentially replacing all of the ideal marginal probability $p_0(z_1...z_{k-1} 0)$ above with its error-mitigated estimator $\hat{p}_{\mathrm{em}}(z_1...z_{k-1} 0)$. In the $k$th step, we will perform a similar hypothesis testing as \cref{eqn:select_hypo}:
\begin{equation}
    \begin{split}\label{eqn:select_hypo_min_str}
    &\hat{p}_{\mathrm{em}}(z_1...z_{k-1}0) \leq p_{\mathrm{th}} \\
    &\quad \Rightarrow \text{Accept Null: } p_{\mathrm{em}}(z_1...z_{k-1}0) = 0 \Rightarrow z_k = 1 ;\\
    &\hat{p}_{\mathrm{em}}(z_1...z_{k-1}0) > p_{\mathrm{th}} \\
    &\quad \Rightarrow \text{Reject Null: } p_{\mathrm{em}}(z_1...z_{k-1}0) > 0 \Rightarrow z_k = 0.
    \end{split}
\end{equation}
The standard deviation of $\hat{p}_{\mathrm{em}}(z_1...z_{k-1} 0)$ can also be obtained via \cref{eqn:var_pem} with $\hat{r}_z$ replaced by the fraction of response circuit runs that has the first $k$ bits of its output being $z_1...z_{k-1} 0$. This standard deviation can then be used to calculate $p_{\mathrm{th}}$ used in the hypothesis testing in the frequentist's approach as discussed before. 

It is also possible to use $p_{\mathrm{th}} = p_{\mathrm{b}}/2$ just before (see \cref{sec:hypo_test_alt}). However, this will on average perform less well than using the same threshold in \cref{sec:sm_str_from_distr}. If we have additional information about the marginal probability $p_{\mathrm{em}}(z_1...z_{k-1}0)$ to construct better $p_{\mathrm{th}}$ for different $k$, then the method in this section has the potential for achieving significant advantages over  \cref{sec:sm_str_from_distr} since it is testing digit-by-digit rather than string-by-string, which means up to an exponential reduction in the number of hypothesis tests needed.

\begin{figure}[htbp]
    \centering
    \subfloat[$10^5$ samples]{\includegraphics[width=0.4\textwidth]{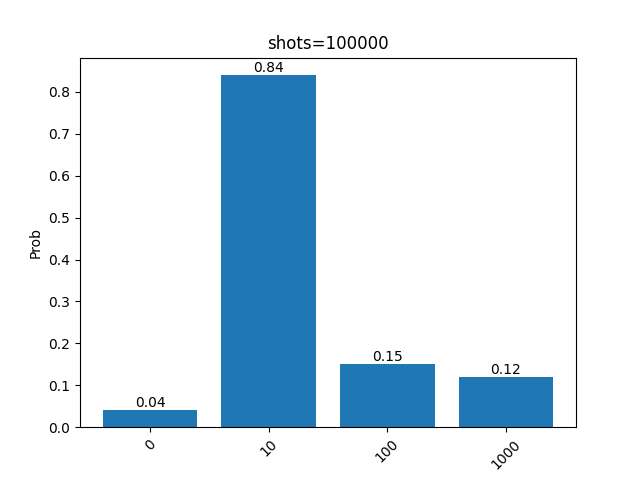}}\\
    \subfloat[$10^6$ samples]{\includegraphics[width=0.4\textwidth]{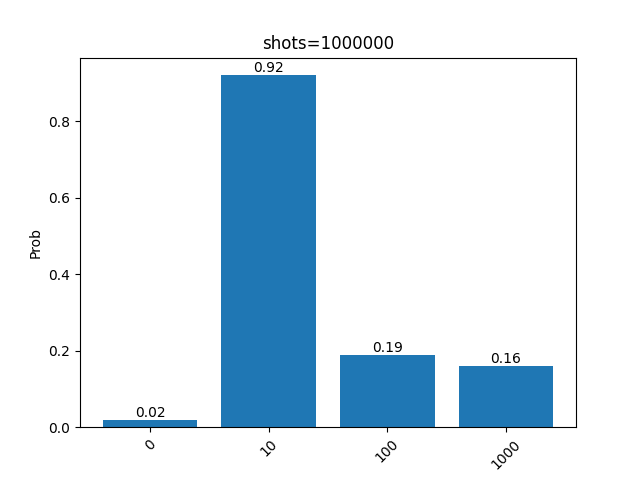}} 
    \caption{Marginal probabilities for recovering the smallest string.}
    \label{fig:alt_method_marginal_prop}
\end{figure}

Using the same numerical example as \cref{sec:numerical_simulation}, the marginal probability estimator $\hat{p}_{\mathrm{em}}(z_1...z_{k-1}0)$ for different digit $k$ is plotted in \cref{fig:alt_method_marginal_prop}. As discussed \cref{sec:numerical_simulation}, the ideal smallest string is $1000$. Hence, the two plots in \cref{fig:alt_method_marginal_prop}, we will need the threshold to be $0.04 \leq p_{\mathrm{th}} \leq 0.12$ and $0.02 \leq p_{\mathrm{th}} \leq 0.16$, respectively to obtain the correct result. 

\section{Thresholds for Hypothesis Testing}
\subsection{Main Method}\label{sec:hyp_test}
\begin{figure}[htbp]
    \centering
    \includegraphics[trim=2cm 2.5cm 2cm 3.5cm, clip, width=\columnwidth]{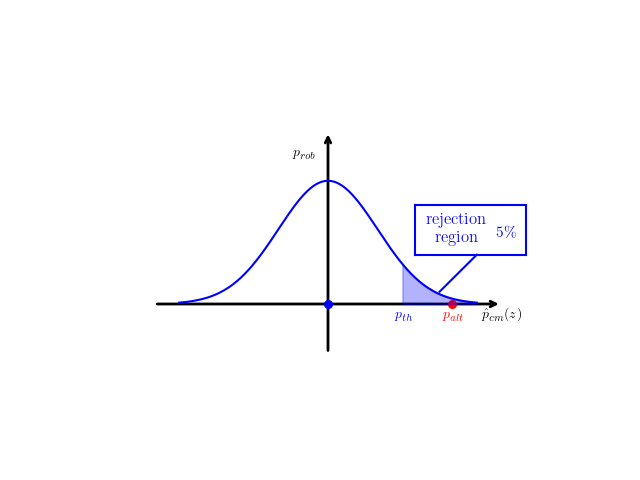}    
    \caption{Threshold for the frequentist's hypothesis testing.}
    \label{fig:freq_threshold}
\end{figure}

In the frequentist's approach, we will only care about bounding the type I error (false rejection of the null hypothesis). If we want to ensure the type I error rate (or $p$-value) is below $0.05$, we need to set the threshold to be
\begin{align*}
    p_{\mathrm{th}}(z) \approx 1.65  \hat{\sigma}[\hat{p}_{\mathrm{em}}(z)] &\approx 1.65 \epsilon\sqrt{\hat{r}_z}.
\end{align*}
This is shown in \cref{fig:freq_threshold}. 

We will start from the smallest sampled string and move up one by one. For the strings that have $\hat{p}_{\mathrm{em}}(z) \leq 0$, we will directly assume that they correspond to $p_{\mathrm{em}}(z) = 0$ and skip them. For the rest of the strings, we will perform the test in \cref{eqn:select_hypo}, and the first string we meet that output $p_{\mathrm{em}}(z) > 0$ will be the smallest string that we found. 

\begin{figure}[htbp]
    \includegraphics[trim=2cm 2.5cm 2cm 3cm, clip, width=\columnwidth]{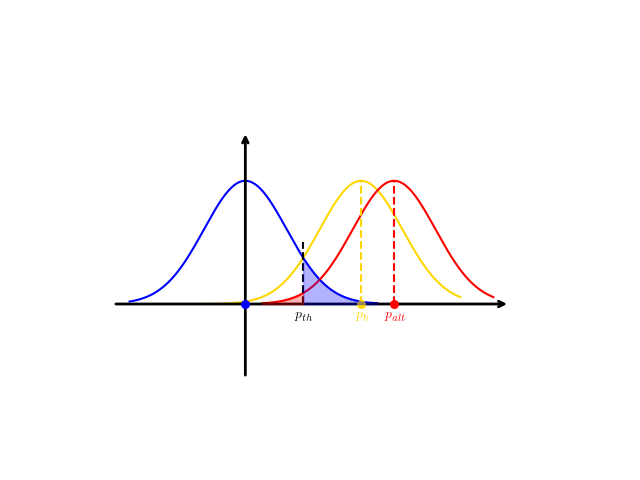}    
    \caption{Threshold for the Bayesian's hypothesis testing.}
    \label{fig:bayes_threshold}
\end{figure}

If we want to consider the full error rate including type II errors (false acceptance of null), we can turn to Bayesian hypothesis testing in which we build a model around the alternative hypothesis as well. In order to do this, we make use of $p_{0}(z_{\mathrm{min}}) \geq p_{\mathrm{b}}$ to lowerbound the $p_{\mathrm{alt}}(z)$ in \cref{eqn:select_hypo} using $p_{\mathrm{alt}}(z) \geq p_{\mathrm{b}}$. Given that $p_{\mathrm{alt}}(z)$ is unknown, we can pick the alternative hypothesis to be $p_{\mathrm{em}}(z) =  p_{\mathrm{b}}$ with a similar standard deviation given by \cref{eqn:var_pem}. This is shown in \cref{fig:bayes_threshold}, in which we see that such an approximation has led to larger type I errors than type II.

\subsection{Alternative Method}\label{sec:hypo_test_alt}
We can still use the frequentist approach with the sampled variance of the estimators. However, as described before, such an approach does not give a bound on the type II errors and thus it is hard to compare the results of this approach against \cref{sec:sm_str_from_distr}. 

If we turn to Bayesian hypothesis testing, we need to construct an alternative hypothesis for the marginal probability $p_{\mathrm{em}}(z_1...z_{k-1} 0)$ beyond the null hypothesis of $p_{\mathrm{em}}(z_1...z_{k-1} 0) = 0$. We know that in the alternative hypothesis, we have $z_k = 0$, and thus $p_{\mathrm{em}}(z_1...z_{k-1} 0) = p_{\mathrm{em}}(z_1...z_{k-1} z_k) \geq p_{\mathrm{em}}(z_1...z_{N}) \geq p_{\mathrm{b}}$. Hence, following the similar arguments before, we have set the lower bound $p_{\mathrm{em}}(z_1...z_{k-1}0) = p_{\mathrm{b}}$ as our alternative hypothesis with the threshold probability being $p_{\mathrm{th}} = p_{\mathrm{b}}/2$. 

This approximation of $p_{\mathrm{em}}(z_1...z_{k-1}0) = p_{\mathrm{b}}$ is on average worse than the approximation we made in the alternative hypothesis in \cref{sec:sm_str_from_distr} (which is $p_{\mathrm{em}}(z_1...z_{N}) = p_{\mathrm{b}}$), especially for small $k$, thus using Bayesian hypothesis testing using a fixed threshold $p_{\mathrm{th}}$ here on average will not outperform the Bayesian hypothesis testing in \cref{sec:sm_str_from_distr}. We say on average here because shot noise can lead to negative probabilities for some entries which means that sometimes the marginal probability can be smaller than the probability of individual entries. We have proven in \cref{sec:comp_pth} that when using a single threshold $p_{\mathrm{th}}$ for testing the hypothesis in all digits, the range of valid values of $p_{\mathrm{th}}$ that output the correct answer here is on average smaller than the range of valid values of $p_{\mathrm{th}}$ in \cref{sec:sm_str_from_distr}. Hence, one will only apply Bayesian hypothesis testing using the method in this section if there is additional information about the marginal probability $p_{\mathrm{em}}(z_1...z_{k-1}0)$ for different $k$ that allow us to construct better alternative hypothesis than $p_{\mathrm{em}}(z_1...z_{k-1}0) = p_{\mathrm{b}}$. 

When assuming a uniform prior, choosing one hypothesis or another essentially becomes the problem of comparing the likelihood function $P\left[\hat{p}_{\mathrm{em}}(z) | p_{\mathrm{em}}(z) = 0\right]$ and  $P\left[\hat{p}_{\mathrm{em}}(z) | p_{\mathrm{em}}(z) = p_{\mathrm{b}}\right]$ between the two hypothesis. Since the two likelihood functions have the same standard deviation obtained from the samples, a balanced strategy that leads to equal type I and type II error rates is to pick a threshold value $p_{\mathrm{th}}(z) = p_{\mathrm{b}}/2$ right between the two means and select the hypothesis using \cref{eqn:select_hypo}. i.e. we have the same threshold for all $z$.

However, note that the true alternative hypothesis should actually be $p_{\mathrm{em}}(z) =  p_{\mathrm{alt}}(z)$. The type II error we obtained by using $p_{\mathrm{em}}(z) = p_{\mathrm{b}}$ as the alternative hypothesis actually gives an upper bound of the actual type II error. Hence, our 'balance' strategies above with $p_{\mathrm{th}}(z) = p_{\mathrm{b}}/2$ actually will have a larger type I error rate than the type II error rate, i.e. it biases towards the null hypothesis of $p_{\mathrm{em}}(z) =  0$. That is to say, when the number of circuit runs is small and the shot noise is big, our estimate will usually lead to an upper bound of the smallest string and is much less likely to be smaller than the true smallest string. 

\subsection{Comparing the threshold probability range}\label{sec:comp_pth}
In order to output the right result using Bayesian hypothesis testing in \cref{sec:sm_str_from_distr} with the smallest string in the distribution being $z_{\mathrm{min}}$, we need to have a threshold value that satisfies:
\begin{equation}
    \begin{split}\label{eqn:pth_cond}
        p_{\mathrm{th}} &< \hat{p}_{\mathrm{em}}(z_{\mathrm{min}})\\
    p_{\mathrm{th}} &\geq \hat{p}_{\mathrm{em}}(z) \quad \forall z < z_{\mathrm{min}}
    \end{split}
\end{equation}

For the rest of this section, we will consider the Bayesian hypothesis testing using the method in \cref{sec:sm_str_alt} with the smallest string in the distribution being $z_{\mathrm{min}} = z_1z_2...z_N$, its null hypothesis is accepted if
\begin{align*}
    \hat{p}_{\mathrm{em}}(z_1...z_{k-1}0) \leq p_{\mathrm{th}} \Rightarrow p_{\mathrm{em}}(z_1...z_{k-1}0) = 0 \Rightarrow z_k = 1
\end{align*}
which implies
\begin{align}\label{eqn:z1_restrict_2}
    p_{\mathrm{em}}(z_1...z_{k}) = 1-p_{\mathrm{em}}(z_1...z_{k-1}0) = 1
\end{align}
and
\begin{align}\label{eqn:z1_restrict}
    z_k = 1 \Rightarrow 1 - \hat{p}_{\mathrm{em}}(z_1...z_{k}) \leq p_{\mathrm{th}}
\end{align}

The alternative hypothesis is accepted (i.e. the null hypothesis is rejected) if 
\begin{align*}
    \hat{p}_{\mathrm{em}}(z_1...z_{k-1}0) > p_{\mathrm{th}} \Rightarrow p_{\mathrm{em}}(z_1...z_{k-1}0) > 0 \Rightarrow z_k = 0
\end{align*}
which implies $p_{\mathrm{em}}(z_1...z_{k}) > 0$ and 
\begin{align}\label{eqn:z0_restrict}
    z_k = 0 \Rightarrow \hat{p}_{\mathrm{em}}(z_1...z_{k}) > p_{\mathrm{th}}
\end{align}
The larger the $k$ is, usually the marginal probability $p_{\mathrm{em}}(z_1...z_{k})$ will be smaller and so is its estimator $\hat{p}_{\mathrm{em}}(z_1...z_{k})$. Hence, this bound in \cref{eqn:z0_restrict} will be tighter by looking at the highest digit of $0$. 

Suppose the last digit of $z_{\mathrm{min}}$ is $z_N = 0$ and $k_1$ is the highest digit of $1$ in $z_{\mathrm{min}}$. Then a necessary and sufficient condition for $p_{\mathrm{th}}$ to give the right result is given by \cref{eqn:z0_restrict} with $k = N$ and \cref{eqn:z1_restrict} with $k=k_1$
\begin{align*}
    \hat{p}_{\mathrm{em}}(z_1...z_{N})  >  p_{\mathrm{th}} \geq 1 - \hat{p}_{\mathrm{em}}(z_1...z_{k_1})
\end{align*}
We know that $k_1$ is the largest digits of $1$, thus any strings that is contained in $\hat{p}_{\mathrm{em}}(z_1...z_{k_1})$ will be larger than $z_{\mathrm{min}}$, which give us
\begin{align*}
    & 1 - \hat{p}_{\mathrm{em}}(z_1...z_{k_1}) \\
    & = 1 - (\hat{p}_{\mathrm{em}}(z_{\mathrm{min}}) + \sum_{\substack{y_{k_1 + 1}... y_{N} \\> 0...0}} \hat{p}_{\mathrm{em}}(z_1...z_{k_1} y_{k_1 + 1}... y_{N}))\\
    & \geq \sum_{z < z_{\mathrm{min}}} \hat{p}_{\mathrm{em}}(z)
\end{align*}
Hence, for $z_N = 0$, 
\begin{align}\label{eqn:pth_cond_alt}
    \hat{p}_{\mathrm{em}}(z_{\mathrm{min}})  >  p_{\mathrm{th}} \geq 1 - \hat{p}_{\mathrm{em}}(z_1...z_{k_1}) \geq \sum_{z < z_{\mathrm{min}}} \hat{p}_{\mathrm{em}}(z)
\end{align}
The upper bound is the same as \cref{eqn:pth_cond}. For the lower bound, we have the sum over all the strings that is smaller than $z_{\mathrm{min}}$, whose variance is also the sum of the variance of individual $\hat{p}_{\mathrm{em}}(z)$. Hence, even though $\sum_{z < z_{\mathrm{min}}} \hat{p}_{\mathrm{em}}(z)$ and individual $\hat{p}_{\mathrm{em}}(z)$ have the same expectation value at $0$, we will still expect in general $\sum_{z < z_{\mathrm{min}}} \hat{p}_{\mathrm{em}}(z) > \hat{p}_{\mathrm{em}}(z)$ due to the larger variance of $\hat{p}_{\mathrm{em}}(z)$. Hence, the lower bound here is higher than \cref{eqn:pth_cond}, which means that the method in \cref{sec:sm_str_alt} has a more restricted range of $p_{\mathrm{th}}$ compared to the method in \cref{sec:sm_str_from_distr} when the last digit of $z_{\mathrm{min}}$ is $z_N = 0$. 

Suppose the last digit is $z_N = 1$ and $k_0$ is the highest digit of $0$ in $z_{\mathrm{min}}$. Then a necessary and sufficient condition for $p_{\mathrm{th}}$ to give the right result is given by \cref{eqn:z0_restrict} with $k = k_0$ and \cref{eqn:z1_restrict} with $k=N$
\begin{align*}
    \hat{p}_{\mathrm{em}}(z_1...z_{k_0})  &>  p_{\mathrm{th}} \geq 1 - \hat{p}_{\mathrm{em}}(z_1...z_{N})\\
    1 - \hat{p}_{\mathrm{em}}(z_1...z_{k_0})  &<  1 - p_{\mathrm{th}} \leq  \hat{p}_{\mathrm{em}}(z_1...z_{N})
\end{align*}
We know that $k_0$ is the largest digits of $0$ in $z_{\mathrm{min}}$, thus any strings that is contained in $\hat{p}_{\mathrm{em}}(z_1...z_{k_0})$ will be smaller than $z_{\mathrm{min}}$, which give us
\begin{align*}
    & 1 - \hat{p}_{\mathrm{em}}(z_1...z_{k_0}) \\
    & = 1 - (\hat{p}_{\mathrm{em}}(z_{\mathrm{min}}) + \sum_{\substack{y_{k_1 + 1}... y_{N} \\< 1...1}} \hat{p}_{\mathrm{em}}(z_1...z_{k_0} y_{k_0 + 1}... y_{N}))\\
    & \geq \sum_{z > z_{\mathrm{min}}} \hat{p}_{\mathrm{em}}(z)
\end{align*}
Hence, for $z_N = 1$, 
\begin{align}
    \hat{p}_{\mathrm{em}}(z_{\mathrm{min}})  \geq  1 - p_{\mathrm{th}} > 1 - \hat{p}_{\mathrm{em}}(z_1...z_{k_0}) \geq \sum_{z > z_{\mathrm{min}}} \hat{p}_{\mathrm{em}}(z)
\end{align}
Note that we have written the inequality in terms of $1 - p_{\mathrm{th}}$ for easier comparison to \cref{eqn:pth_cond_alt} and \cref{eqn:pth_cond}. The upper bound for $1 - p_{\mathrm{th}}$ here is the same as the upper bound for $p_{\mathrm{th}}$ in \cref{eqn:pth_cond_alt}. The lower bound however is in general higher than that of \cref{eqn:pth_cond_alt} since the expectation value of $\sum_{z > z_{\mathrm{min}}} \hat{p}_{\mathrm{em}}(z)$ is not zero unless the whole distribution only has a single non-zero entry at $z_{\mathrm{min}}$. Hence, the interval of value that can be taken by $1 - p_{\mathrm{th}}$ (and thus $p_{\mathrm{th}}$) when $z_N = 1$ is smaller than that in \cref{eqn:pth_cond_alt} when  $z_N = 0$, and they are both smaller than the $p_{\mathrm{th}}$ interval in \cref{eqn:pth_cond} when using the method in \cref{sec:sm_str_from_distr} instead. 

\section{Application in Eigenstate Preparation}\label{sec:eigenstate}
Given a starting state $\ket{\psi}_{\mathrm{in}} = \sum_{i = 0} \sqrt{p_i}\ket{\phi_i}$ where $\ket{\phi_i}$ are the eigenstates of some given Hamiltonian. One way to prepare the ground state $\ket{\phi_0}$ is to perform quantum phase estimation (QPE) on $\ket{\psi}_{\mathrm{in}}$, turning it into the state $\ket{\psi} = \sum_{i = 0} \sqrt{p_i} \ket{E_i}\ket{\phi_i}$ and then measuring the energy register to post-select on the ground state energy $E_0$. The output state will be the target ground state $\ket{\phi_0}$, using which we can measure any target observable $O$ we want. In other words, we can obtain the ground state energy $E_0$ by running QPE many times and selecting the smallest string. After that, the ground state property of some observable $O$ can be obtained by simply measuring $\ketbra{E_0} \otimes O$ on the $\ket{\psi}$ state after the QPE circuit. 

Moving to the noisy case, the error-mitigated value of $E_0$ can simply be obtained via the method outlined in this section. After that, the main problem of measuring the expectation value of $\ketbra{E_0} \otimes O$ on the output state corrupted by noise is naturally compatible with all the QEM techniques. We can use the same method to prepare other eigenstates as long as the target eigenstate has sufficient overlap with the input state $\ket{\psi}_{\mathrm{in}}$.

\section{Further Numerics}\label{sec:further_numerics}
In the simulation, we have modified the example in \cite{QiskittextbookContentChalgorithms} by flipping some bits of the energy value of all the eigenstates for a clearer illustration of our results, which is equivalent to a universal shift in the absolute value of the energy.  

\subsection{Different number of circuit runs}
\begin{figure}[htbp]
    \includegraphics[width=0.475\textwidth ]{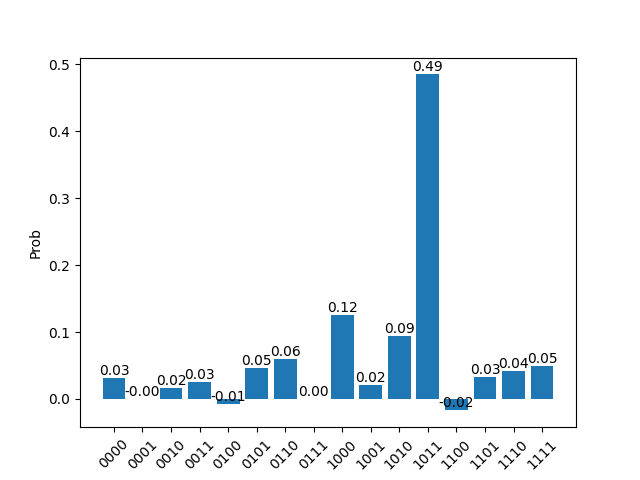}
    \caption{Error-mitigated distribution with $10^5$ circuit runs.}
    \label{fig:mitigated_distr_less_samp}
\end{figure}

Reducing the number of circuit runs to $10^5$ for the example in the main text will output \cref{fig:mitigated_distr_less_samp}, which still has a very high resemblance to the ideal distribution with a total square error of $\mathrm{TSE}_{\mathrm{em}} = {0.056}$. This error is unsurprisingly higher than the $0.004$ in the $10^6$-run case, but still much lower than $0.297$ in the noisy case. The range of the threshold we can take to obtain the correct minimum string from the distribution is now given by $0.06 \leq p_{\mathrm{th}} < 0.12$, which is also a reduced range compared to the $10^6$-run case. 

\subsection{Different number of qubits}
\begin{figure}[hbtp]
    \centering
        \subfloat[Ideal, $p_0(z)$]{\includegraphics[width=0.47\textwidth]{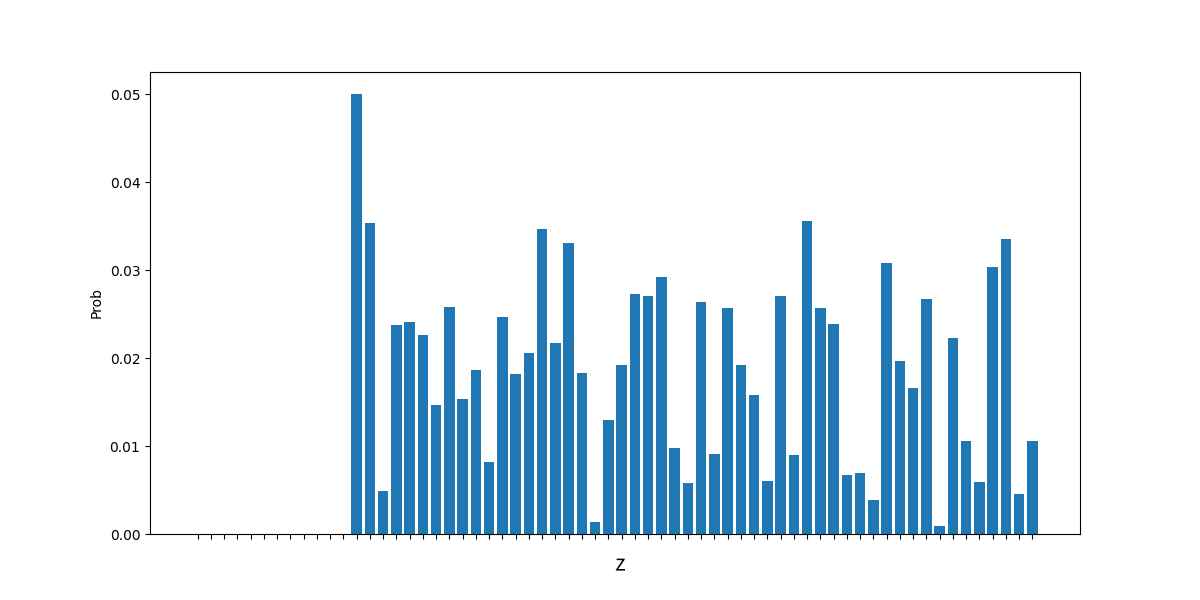}} \\
        \subfloat[Noisy, $p(z)$]{\includegraphics[width=0.47\textwidth]{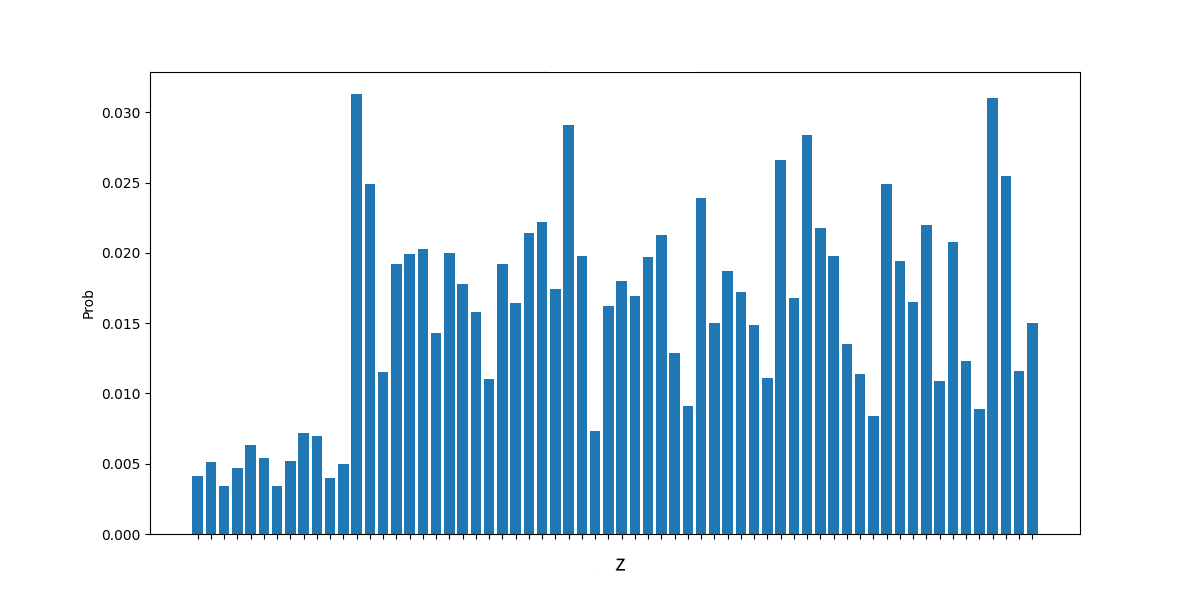}} \\
        \subfloat[Mitigated,  $\hat{p}_{\mathrm{em}}(z)$]{\includegraphics[width=0.47\textwidth]{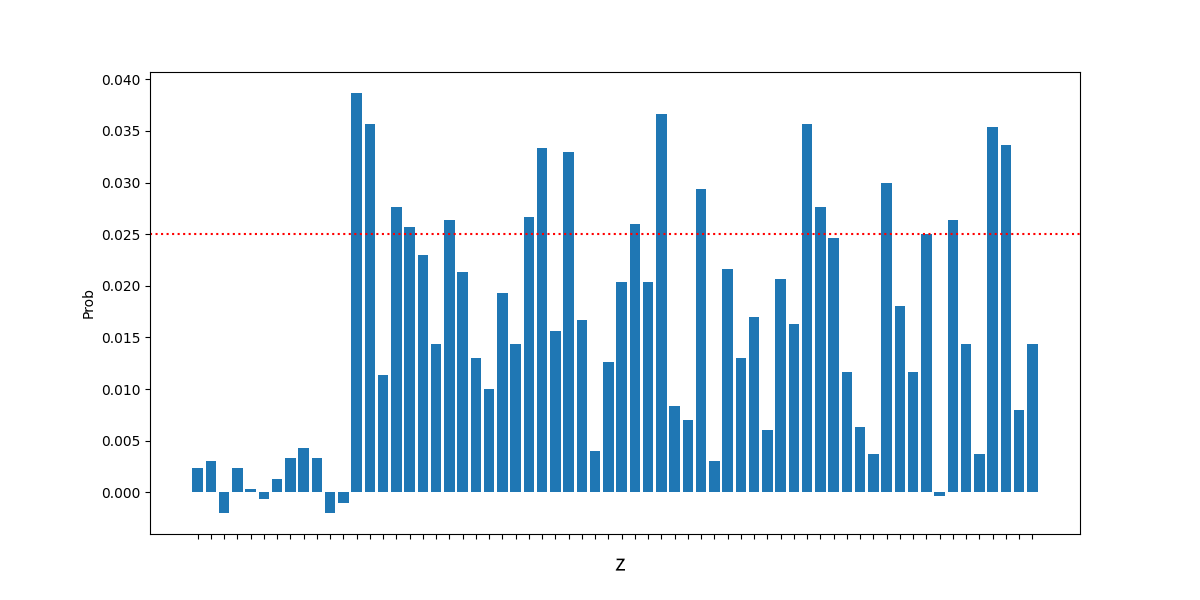}} 
        \caption{Output distribution of 6-qubit simulation. 
        }
        \label{fig:6_qubits}
\end{figure}

It is resource-intensive to perform noisy simulations of qubit numbers beyond the main text due to the limited amount of computing resources we have access to. Instead, we will focus on measurement noise and see how our schemes can be used to mitigate output distributions affected by measurement noise. The level of noise is kept at the similar $\order{1}$ number of faults per circuit run mentioned in the main text. For $6$-qubit example, the result is shown in \cref{fig:6_qubits}. Similar to before, the error-mitigated distribution is much closer to the ideal distribution compared to the noisy one.

\newpage
\bibliographystyle{apsrev4-2}
%

\end{document}